\documentclass[aps,prd,showpacs,showkeys,preprint,nofootinbib]{revtex4}
\usepackage[british]{babel}

\usepackage{amsmath}
\usepackage{amssymb}

\usepackage{bbm}

\DeclareFontFamily{OT1}{rsfs}{}
\DeclareFontShape{OT1}{rsfs}{m}{n}{ <-7> rsfs5 <7-10> rsfs7 <10->rsfs10}{} 
\DeclareMathAlphabet{\mycal}{OT1}{rsfs}{m}{n}
\newcommand{\lan}{{\mycal L}}

\newcommand{\p}{\partial}

\newcommand{\vp}{\mathbf{p}}

\newcommand{\dual}[1]{\overset{{}^{{}^{\boldsymbol{\neg}}}}{\smash[t]{#1}}}

\begin{document}

\title{The Einstein-Cartan-Elko system}

\author{C. G. B\"ohmer}
\email{boehmer@hep.itp.tuwien.ac.at}
\affiliation{ASGBG/CIU, Department of Mathematics, Apartado Postal C-600, 
             University of Zacatecas (UAZ), Zacatecas, Zac 98060, Mexico}

\date{21. July 2006}

\begin{abstract}
The present paper analyses the Einstein-Cartan theory of gravitation with
Elko spinors as sources of curvature and torsion. After minimally
coupling the Elko spinors to torsion, the spin angular momentum tensor is
derived and its structure is discussed. It shows a much richer structure 
than the Dirac analogue and hence it is demonstrated that spin one half
particles do not necessarily yield only an axial vector torsion component.
Moreover, it is argued that the presence of Elko spinors partially solves
the problem of minimally coupling Maxwell fields to Einstein-Cartan theory.
\end{abstract}


\maketitle

\section{Introduction}

Einstein-Cartan theory is probably the simplest and straightest generalisation
of Einstein's theory of general relativity. It is based on the usual 
Einstein-Hilbert action, however, it is not assumed that torsion
vanishes, as was originally done by Einstein. If in the Einstein-Hilbert
action the metric and the torsion are considered as independent variables,
then the variations with respect to them yield two field equations. The
first one relates the (non-symmetric) Einstein tensor to the canonical
energy-momentum tensor, whereas the second one relates torsion to the
spin angular momentum of matter, see e.g.~\cite{Trautman:1972aa,Trautman:1973aa,Hehl:1973,Hehl:1976kj,Hehl:1995ue,Hammond:2002rm}.

In the usual Einstein gravity, matter (mass) couples to the curvature of 
spacetime, whereas spin does not couple to geometrical quantities. 
The Poincar\'e Lie algebra can however be classified by the values of 
the two Casimir operators, see e.g.~\cite{waldgeneral}, mass squared $M^2$ and angular 
momentum squared $S^2=s(s+1)$, where $s$ is the spin taking the usual values 
$0,\pm 1/2,\ldots$ Therefore it seems quite natural~\cite{Hehl:1976kj,Hammond:2002rm} 
to consider a theory of gravitation that takes both quantities, 
i.e.~{\em mass and spin}, into account. The minimal coupling of standard
model matter fields has been thoroughly discussed in~\cite{Hehl:1976kj}
and the physical consequences were analysed. Ever since it allegedly became
well known that spin one half matter fields only couple to the 
axial vector part of the torsion tensor, and that higher spin massive
particles couple also to other parts of the torsion tensor~\cite{Garcia:1998jw}.
It turns out, however, that the spin one half Elko spinors do couple
to all parts of the torsion tensor, although they are spin one half.
To understand these new properties from a physical point of view, some
facts about the recently discovered Elko spinors should be recalled~\cite{jcap}, 
with a condensed version published in~\cite{prd}. 
The Elko spinors belong to a wider class of so-called flagpole spinors, 
see~\cite{daRocha:2005ti} for the Lounesto spinor field classification.
The term Elko spinors originates from the German
{\itshape {\bfseries E}igenspinoren des {\bfseries L}adungs{\bfseries k}onjugations{\bfseries o}perators} (eigenspinors of the charge conjugation operator).

Elko spinors are based on the eigenspinors of the charge conjugation operator,
are non-standard Wigner class spinors and therefore yield a non-local field
theory. They obey the unusual property $(CPT)^2=-\mathbbm{1}$. More 
explicitely, an Elko spinor is defined by~\cite{jcap,prd}
\begin{align}
      \lambda = \begin{pmatrix} \pm \sigma_2 \phi^{\ast}_{L} \\
                \phi_L \end{pmatrix} \,,
      \label{eq:i1}
\end{align}
where $\phi^{\ast}_{L}$ denotes the complex conjugate of $\phi_L$
and $\sigma_2$ denotes the second Pauli matrix.
The upper sign stands for the self conjugate spinor and the lower sign
for the anti self conjugate spinor with respect to the charge
conjugation operator. The same physical content would have been
obtained, had one started with the right-handed two-spinor $\phi_{R}$.
For all details regarding the field theory of Elko spinors I refer the 
reader to the two fundamental papers~\cite{jcap,prd}. The two two-spinors
$\sigma_2 \phi^{\ast}_{L}$ and $\phi_L$ have opposite helicities and
hence should be distinguishable. To do so, one writes
 \begin{align}
      \lambda_{\{-,+\}} = \begin{pmatrix} \pm \sigma_2 \phi^{+}_{L}{}^{\ast} \\
                \phi^{+}_L \end{pmatrix} \,, \qquad
      \lambda_{\{+,-\}} = \begin{pmatrix} \pm \sigma_2 \phi^{-}_{L}{}^{\ast} \\
                \phi^{-}_L \end{pmatrix} \,,
      \label{eq:i2}
\end{align}
where the first entry of the helicity subscript $\{-,+\}$ refers to the upper 
two-spinor and the second the lower two-spinor. This subscript is henceforth
denoted by indices $u,v,\ldots$ Since the Elko spinors with respect to the
standard Dirac dual $\bar{\psi} = \psi^{\dagger}\gamma^0$ have an 
imaginary bi-orthogonal norm~\cite{jcap,prd}, a new dual has to be defined
so that a consistent field theory emerges. This Elko dual is given by
\begin{align}
       \dual{\lambda}_u = i\,\varepsilon_u^v \lambda_v^{\dagger} \gamma^0 \,,
       \label{eq:i3}
\end{align}
with the skew-symmetric symbol 
$\varepsilon_{\{+,-\}}^{\{-,+\}}=-1=-\varepsilon_{\{-,+\}}^{\{+,-\}}$ (note, 
that for Dirac spinors the term $i\,\varepsilon_u^v$ is just $\delta_u^v$). 
With the dual defined above one finds (by construction) the standard relation
\begin{align}
      \dual{\lambda}_u(\vp) \lambda_v(\vp) = \pm\, 2m\, \delta_{uv} \,,
      \label{eq:i4}
\end{align}
where $\vp$ denotes the momentum.

As it has been pointed out above, the Elko spinors have a double helicity
structure, opposed to Dirac spinors, where both two-spinors have the
same helicity. The key feature of the Einstein-Cartan theory is, that spin
couples to certain parts of geometry, i.e.~torsion. Therefore, it is clear 
that this double helicity structure is much more sensitive to torsion than 
in the analogue Dirac case. Hence it is also expected (and shown in 
Section~\ref{sec:elko}) that the minimal coupling of Elko spinors to torsion 
yields a more interesting torsion tensor than in the Dirac spinor case. 

The paper is organised as follows. In Section~\ref{sec:spinors} the 
introduction of spinors into spacetimes is recalled. Section~\ref{sec:elko}
describes general aspects of the Einstein-Cartan-Elko system and in
Section~\ref{sec:gauge} gauge couplings in Einstein-Cartan theory
are discussed. Conclusions of this work are presented in the final
Section~\ref{sec:sum}.

\section{Einstein-Cartan theory and spinors}
\label{sec:spinors}

The main aim of this section is to shortly recall some features of the
Einstein-Cartan theory of gravitation, with a particular focus on the
anholonomic formulation. If the starting point is a special relativistic 
field theory, one can apply the ``comma to semicolon'' rule 
$(\p_a \rightarrow \nabla_a)$ to its Lagrangian in order to find the
minimally coupled spacetime field theory.  

Torsion is most naturally taken into account by assuming the existence
of a covariant derivative operator $\tilde{\nabla}_a$ that is not torsion 
free~\cite{Trautman:1972aa,Trautman:1973aa,Hehl:1973,Hehl:1976kj,Hehl:1995ue}. 
Quantities denoted with a tilde always take torsion into account. Therefore the 
minimal scheme to introduce torsion can symbolically by written as 
$\nabla_a \rightarrow \tilde{\nabla}_a$ and the complete path from a special 
relativistic to an Einstein-Cartan field theory can be formulated as
\begin{align}
      \p_a \quad \rightarrow \quad \nabla_a \quad \rightarrow 
      \quad \tilde{\nabla}_a
      \label{eq:rule}
\end{align}
The latter non-torsion free covariant derivative can be split according to
\begin{align}
      \tilde{\nabla}_a \lambda &= \partial_a \lambda - \frac{1}{4} \Gamma_{abc}
      \gamma^b \gamma^c \lambda + \frac{1}{4} K_{abc} \gamma^b \gamma^c \lambda \,,
\label{eq:t1}
\end{align}
where $K_{abc}$ is the contortion (and not contorsion) tensor, 
\begin{align}
      \tilde{\Gamma}^a_{bc} = \Gamma^a_{bc} - K_{bc}{}^{a} \,. 
      \label{eq:t1a}
\end{align}
The skew-symmetric part of the connection defines the torsion tensor 
$S_{bc}{}^{a}$ to be
\begin{align}
      S_{bc}{}^{a} = \tilde{\Gamma}^a_{[bc]} =
      \frac{1}{2} \bigl(\tilde{\Gamma}^a_{bc}-
      \tilde{\Gamma}^a_{cb} \bigr) \,.
      \label{eq:t1b} 
\end{align}
By virtue of the last two equations, torsion and contortion are related by
\begin{align}
      S_{bc}{}^{a} = \frac{1}{2} \bigl(K_{cb}{}^{a}-K_{bc}{}^{a} \bigr) \,,
      \label{eq:t1c}
\end{align}
Denoting the torsion free covariant derivative by $\nabla_a$, the covariant
derivative, when acting on a spinor~(\ref{eq:t1}) can be written in the 
following form  
\begin{align}
      \tilde{\nabla}_a \lambda = \nabla_a \lambda + \frac{1}{4} K_{abc} 
      \gamma^b \gamma^c \lambda \,.
      \label{eq:t2}
\end{align} 
Since the covariant derivative obeys the Leibnitz rule, and the
action of $\tilde{\nabla}_a$ on the scalar $\dual{\lambda}\lambda$
is known, for the dual Elko spinor $\dual{\lambda}$ (see equation~(\ref{eq:i3}))
one therefore finds 
\begin{align}
      \tilde{\nabla}_a \dual{\lambda} = \nabla_a \dual{\lambda} -  
      \frac{1}{4} K_{abc} \dual{\lambda} \gamma^b \gamma^c \,.
      \label{eq:t3}
\end{align}
After having defined, how to introduce a classical field theory into
the Einstein-Cartan theory, one can now formulate the action.
Since the Einstein-Cartan Lagrangian is simply the Einstein-Hilbert
action (with metric and torsion regarded as independent variables),
the coupled field equations can be derived from the total action
given by
\begin{align}
      S = \int \Bigl( \frac{1}{2\kappa} \tilde{R} + \tilde{\lan}_{\rm mat} \Bigr)
      \sqrt{-g}\, d^4 x \,,
      \label{eq:a1}
\end{align}
where the speed of light was set to one $(c=1)$. The Ricci scalar is denoted 
by $\tilde{R}$ (computed from the complete connection with the contortion 
contributions), $g$ is the determinant of the metric, $\tilde{\lan}_{\rm mat}$ 
the (minimally coupled) matter Lagrangian and $\kappa=8\pi G$ is 
the coupling constant. The field equations of Einstein-Cartan 
theory~\cite{Hehl:1976kj} are obtained by varying the total action 
function~(\ref{eq:a1}) with respect to the metric $\eta^{ij}$ and 
the contortion $K^{i}{}_{jk}$ (or torsion) as independent variables, 
which yields
\begin{align}
      \tilde{R}_{ij} - \frac{1}{2} \tilde{R} \eta_{ij} &= \kappa\,\Sigma_{ij} \,,
      \label{eq:t8} \\
      S^{ij}{}_{k} + \delta^i_k S^{j}{}_{l}{}^{l} - \delta^j_k S^{i}{}_{l}{}^{l}
      &= \kappa\,\tau^{ij}{}_{k} \,.
      \label{eq:t9}
\end{align}
In equation~(\ref{eq:t9}) $\tau^{ij}{}_{k}$ is the spin angular momentum tensor
and $\Sigma_{ij}$ in~(\ref{eq:t8}) is the total (or canonical) energy-momentum 
tensor which is not symmetric. It is defined by
\begin{align}
      \Sigma_{ij} = \sigma_{ij} + (\tilde{\nabla}_k+K_{lk}{}^{l})
      (\tau_{ij}{}^{k}-\tau_{j}{}^{k}{}_{i}+\tau^{k}{}_{ij}) \,,
      \label{eq:t10}
\end{align}
where $\sigma_{ij}$ is the metric energy-momentum tensor given by
the variation of the matter Lagrangian with respect to the metric.
It should be emphasised that the trajectories of test particles 
are obtained by integrating the conservation equations of the energy 
and the angular momentum, see e.g.~\cite{Hehl:1976kj}, and in general 
are neither geodesics nor autoparallels. It is also important to 
note that the field equations~(\ref{eq:t9}) are
algebraic equations for the torsion. Therefore $\tau_{ijk}=0$
immediately implies the vanishing of the torsion. Hence torsion
is only present in spacetime regions with torsion sources and
therefore it is not propagating. Because of the algebraic torsion 
equation one can formally eliminate the torsion by its spin sources 
in the action. This yields one effective or combined field equation
\begin{align}
      G_{ij} = \kappa\,\hat{\sigma}_{ij} \,,
      \label{eq:a2}
\end{align}
where the energy-momentum tensor on the right-hand side is given by
the following rather complicated expression~\cite{Hehl:1976kj}
\begin{align}
      \hat{\sigma}_{ij} = \sigma_{ij} + \kappa \Bigl( 
      -4 \tau_{i}{}^{k}_{[l} \tau_{j}{}^{l}{}_{k]} -2\tau_{i}{}^{kl}\tau_{jkl} + 
      +\tau^{kl}{}_{i}\tau_{klj} +\frac{1}{2} \eta_{ij} \bigl(
      4\tau_{m}{}^{k}{}_{[l} \tau^{ml}{}_{k]}+\tau^{klm}\tau_{klm}\bigr)
      \Bigr) \,.
      \label{eq:a3}
\end{align}
Note that the square of the coupling constant $\kappa^2$ enters the torsion 
contributions in $\hat{\sigma}_{ij}$ in equation~(\ref{eq:a3}).
This effective or combined energy-momentum tensor is symmetric and satisfies
the usual conservation equation $\nabla^j \hat{\sigma}_{ij}=0$. 
After recalling some basic principles of Einstein-Cartan theory, the 
minimal coupling of Elko spinors to Einstein-Cartan theory is now
considered in the next section.

\section{Elko spinors in Einstein-Cartan theory}
\label{sec:elko}

The Elko spinors obey scalar field-like equations of motion since their
mass dimension is one. The Elko Lagrangian constructed in~\cite{jcap} reads
\begin{align}
      \lan_{\rm Elko} = \eta^{ab}\p_a \dual{\lambda} \p_b \lambda 
      -m^2 \dual{\lambda} \lambda + \alpha \bigl[\dual{\lambda} \lambda \bigr]^2 \,.
      \label{eq:t}
\end{align}
As described above, one can apply the scheme~(\ref{eq:rule}) to the above 
Lagrangian and arrives at
\begin{align}
      \tilde{\lan}_{\rm Elko} = \eta^{ab}\tilde{\nabla}_a 
      \dual{\lambda} \tilde{\nabla}_b \lambda -m^2 \dual{\lambda} \lambda + 
      \alpha \bigl[\dual{\lambda} \lambda \bigr]^2 \,.
      \label{eq:t0}
\end{align}
One should be slightly careful with the Lagrangian~(\ref{eq:t0}), since if varied
with respect to the metric, the resulting term $\nabla_a \dual{\lambda} \nabla_b \lambda$
is not necessarily symmetric, even in spacetimes without torsion. 
Hence one should add the Elko conjugate equation so that consistent field 
equations emerge. Hence the correct covariant Elko Lagrangian reads
\begin{align}
      \tilde{\lan}_{\rm Elko} = \eta^{ab} \frac{1}{2}\Bigl(\tilde{\nabla}_a \dual{\lambda} 
      \tilde{\nabla}_b \lambda + \tilde{\nabla}_b \dual{\lambda} 
      \tilde{\nabla}_a \lambda \Bigr) -m^2 \dual{\lambda} \lambda + 
      \alpha \bigl[\dual{\lambda} \lambda \bigr]^2 \,.
      \label{eq:t0a}
\end{align}
It should be noted the the Lagrangian of complex massive scalars is formally
very similar to the Elko Lagrangian. The difference, which is in fact crucial,
is that the Elkos are spinors. The covariant derivative, when acting on scalars, 
is just the partial derivative $\tilde{\nabla}_a\phi=\p_a\phi$. Hence, scalar 
particles (spin 0) cannot couple minimally to the torsion of 
spacetime~\cite{Hehl:1976kj}, however, higher spin matter can. 

By taking into account equations~(\ref{eq:t2}) and~(\ref{eq:t3}) one can
split the Elko Lagrangian into its torsion free part and the torsion
contributions and arrives at
\begin{align}
      \tilde{\lan}_{\rm Elko} = \lan_{\rm Elko}
      + \frac{1}{4}K^{a}{}_{bc} \nabla_a \dual{\lambda} \gamma^b \gamma^c \lambda
      - \frac{1}{4}K^{a}{}_{bc} \dual{\lambda} \gamma^b \gamma^c \nabla_a \lambda
      - \frac{1}{16} K^{a}{}_{bc} K_{ade} \dual{\lambda} \gamma^b \gamma^c
      \gamma^d \gamma^e \lambda \,.
      \label{eq:t5}
\end{align}
Variation with respect to the contortion tensor $K^{i}{}_{jk}$ yields the spin angular 
momentum tensor 
\begin{align}
      \tau^{kj}{}_{i} = \frac{\delta \tilde{\lan}_{\rm Elko}}{\delta K^{i}{}_{jk}} =
      \frac{1}{4} \tilde{\nabla}_i \dual{\lambda} \gamma^j \gamma^k \lambda -
      \frac{1}{4} \dual{\lambda} \gamma^j \gamma^k \tilde{\nabla}_i \lambda \,,
      \label{eq:t5a}
\end{align}
which equivalently can be written as follows
\begin{align}
      \tau^{kj}{}_{i}= \frac{1}{4} \nabla_i \dual{\lambda} \gamma^j \gamma^k \lambda -
      \frac{1}{4} \dual{\lambda} \gamma^j \gamma^k \nabla_i \lambda -
      \frac{1}{16} K_{iab} \dual{\lambda} \gamma^j \gamma^k \gamma^a \gamma^b \lambda -
      \frac{1}{16} K_{iab} \dual{\lambda} \gamma^a \gamma^b \gamma^j \gamma^k \lambda \,.
      \label{eq:t6}
\end{align}
An important feature of this spin angular momentum tensor is that is depends
on the contortion (or torsion) itself, and it is obviously not totally
skew-symmetric (and cannot be expressed entirely as an axial torsion vector). 
This fact should indeed be emphasised, Elko spinors are possible sources for 
all irreducible parts of the torsion tensor. In fact, many references on 
Einstein-Cartan theory state that it is well known that spin one half particles 
yield a axial torsion vector. However, as was just shown, this statement
is not completely correct. The non-standard Wigner class Elko spinors (that
are spin one half) lead to more torsion structure than the Dirac spinors. 
Henceforth, the above statement should be formulated more carefully.

In order to be complete, the Elko Lagrangian~(\ref{eq:t0}) implies the 
following metric energy-momentum tensor
\begin{align}
      \sigma_{ij} = \frac{1}{2}\bigl(\tilde{\nabla}_{i} \dual{\lambda} 
      \tilde{\nabla}_{j} \lambda + \tilde{\nabla}_{j} \dual{\lambda} 
      \tilde{\nabla}_{i} \lambda \bigr) - \eta_{ij} \tilde{\lan}_{\rm elko} \,.
      \label{eq:t6em}
\end{align}

In order to value the particular form of the spin angular momentum tensor~(\ref{eq:t6})
for Elko spinors, its form should be compared with those from other field theories.
For sake of completeness, let us note again, that for scalar fields (spin 0) one has
\begin{align}
      \tau_{ijk} = 0 \,,
\end{align} 
while the spin angular momentum tensor of Dirac fermions (spin $1/2$) is given by
\begin{align}
      \tau^{ijk} = \tau^{[ijk]} = \frac{1}{4}\bar{\psi} \gamma^{[i}
      \gamma^j \gamma^{k]} \psi \,,
      \label{eq:t6a}
\end{align}
which is totally skew-symmetric and has only four independent 
components~\cite{Hehl:1976kj}. The next logical field to consider would be 
the Maxwell field, i.e.~the massless spin one gauge field. However, gauge fields 
cannot be minimally coupled to torsion since the resulting terms containing the 
contortion (or torsion) turn out not to be gauge invariant~\cite{Hehl:1976kj}. 
This subtle problem is avoided by considering massive spin one fields. However, 
by assuming a particular form of the torsion tensor and a modified gauge 
transformation, a minimal coupling scheme could still be developed 
in~\cite{Hojman:1978yz}. The same holds for any Yang-Mills type field 
theory, the minimal coupled scheme would yield terms that brake gauge 
invariance. 

In the introduction it was argued that the presence of two Casimir operators, 
mass and spin, serves as an argument why any theory of gravity should take both 
into account. Consequently, Yang-Mills type field theories should also be treated 
using the minimal coupling scheme. From my point of view, to exclude gauge fields 
is as unnatural as excluding spin from any theory of gravitation. This issue 
certainly requires further investigation. 

Let us come back to the spin angular momentum tensor of massive spin one 
fields~\cite{Hehl:1976kj}, which reads
\begin{align}
      \tau_{kji} = \frac{1}{2} \Bigl(
      U_k \tilde{\nabla}_{[j} U_{i]} - U_j \tilde{\nabla}_{[k} U_{i]} \Bigl)
      = \frac{1}{2} \Bigl( U_k \nabla_{[j} U_{i]} - U_j \nabla_{[k} U_{i]} \Bigl)
      + S_{i[j}{}^{l} U_{k]} U_l \,.
      \label{eq:t6b} 
\end{align}
In that latter case the spin angular momentum tensor also depends on the
contortion, like in the Elko spinor case. 

By repeated application of $\gamma^a \gamma^b = 2\eta^{ab} - \gamma^b \gamma^a$
one can change the order of the $\gamma$-matrices in the third term of the
spin angular momentum tensor~(\ref{eq:t6}), 
$\gamma^j \gamma^k \gamma^a \gamma^b \rightarrow \gamma^a \gamma^b \gamma^j \gamma^k$,
so that the last term appears twice and additional terms are are also present.
Hence the spin angular momentum tensor reads
\begin{multline}
      \tau^{kj}{}_{i} = \frac{1}{4} \nabla_i \dual{\lambda} \gamma^j \gamma^k \lambda -
      \frac{1}{4} \dual{\lambda} \gamma^j \gamma^k \nabla_i \lambda -
      \frac{1}{8} K_{iab} \dual{\lambda} \gamma^a \gamma^b \gamma^j \gamma^k \lambda \\+
      \frac{1}{2} K_{i}{}^{jk} \dual{\lambda} \lambda + \frac{1}{4} K_{ia}{}^{j} 
      \dual{\lambda} \gamma^a \gamma^k \lambda -
      \frac{1}{4} K_{ia}{}^{k} \dual{\lambda} \gamma^a \gamma^j \lambda\,.
      \label{eq:t7}
\end{multline}

With the help of equation~(\ref{eq:t1c}) one can rewrite the second field 
equation~(\ref{eq:t9}) in terms of the contortion tensor
\begin{align}
      K^{ji}{}_{k}-K^{ij}{}_{k}+\delta^i_k K_{l}{}^{jl}-\delta^j_k K_{l}{}^{il} =
      2 \kappa\, \tau^{ij}{}_{k}
      \label{eq:t11}
\end{align}
Inserting in the latter equation $\tau^{ij}{}_{k}$ as given by~(\ref{eq:t7}) leads to
\begin{multline}
      K^{ji}{}_{k} - K^{ij}{}_{k} + \delta^i_k K_{l}{}^{jl} - \delta^j_k K_{l}{}^{il} =
      k \Bigl(\frac{1}{2} \nabla_k \dual{\lambda} \gamma^j \gamma^i \lambda -
      \frac{1}{2} \dual{\lambda} \gamma^j \gamma^i \nabla_k \lambda \\
      - \frac{1}{4} K_{kab} \dual{\lambda} \gamma^a \gamma^b \gamma^j \gamma^i \lambda +
      K_{k}{}^{ji} \dual{\lambda} \lambda + \frac{1}{2} K_{ka}{}^{j} 
      \dual{\lambda} \gamma^a \gamma^i \lambda -
      \frac{1}{2} K_{ka}{}^{i} \dual{\lambda} \gamma^a \gamma^j \lambda
      \Bigr) \,.
      \label{eq:t12}
\end{multline}
The latter equation is an algebraic relation for the 24 component of the contortion
tensor, that can in principle be solved 
$K = K(\dual{\lambda},\lambda,\nabla\dual{\lambda},\nabla\lambda)$. The Elko
spinors appear quadratically in the contortion tensor $K_{bc}{}^{a}$. Because
of the algebraic equation of motion for contortion one can eliminate the contortion
tensor in the Elko Lagrangian. Since the last term in the Elko Lagrangian~(\ref{eq:t5})
is quadratic in the contortion tensor this yields an effectively sextic Elko
self-interaction term. This is quite similar to what happens in the Dirac spinor
case. In that case the Lagrangian is linear in the contortion, and the equations of motion
yield the contortion quadratically in the spinors so that the Lagrangian, after
eliminating the contortion, contains a quartic self-interaction term.

One could furthermore decompose the contortion tensor into its irreducible part, 
the contortion vector $K_{b}=K_{ab}{}^{a}$, the axial vector 
$A_a = \varepsilon_{abcd} K^{bcd}$ and a tensor $Q_{abc}$, say, containing
the remaining contortion parts. In case of Dirac spinors, see~(\ref{eq:t6a}),
this turns out to be quite useful. For Elko spinors however, this is not very
enlightening, since all components essentially enter equation~(\ref{eq:t12}). 

\section{Elko spinors and gauge couplings}
\label{sec:gauge}

As was already pointed out in the previous section, gauge fields cannot be
consistently coupled to Einstein-Cartan theory, since the resulting term
would spoil gauge invariance. Ways out have been proposed, e.g.~in 
Ref.~\cite{Hojman:1978yz} by assuming a rather special form of the torsion
tensor. In what follows it will be argued that the existence of Elko spinors 
in nature might partially solve this problem. The dominant interaction
of Elko spinors~\cite{jcap} are with gravity~(\ref{eq:t0}) and through 
the Higgs doublet with the following interaction Lagrangian
\begin{align}
      \lan_{\rm int}=c_{\rm H}\, \Phi^{\dagger} \Phi\, \dual{\lambda} \lambda \,,
      \label{eq:g1}
\end{align}
where $\Phi$ denotes the Higgs doublet. However, one can also construct
an interaction Lagrangian where the Elko spinor couples to an Abelian
gauge field~\cite{jcap} with field strength $F_{ab}$ that takes the form
\begin{align}
      \lan_{\rm int}=c_{\rm F}\, \dual{\lambda} [\gamma^a,\gamma^b] F_{ab}\, \lambda \,.
      \label{eq:q2}
\end{align}
Such an interaction yields effective mass terms for the photon, so that the
coupling constant $c_{\rm F}$ is heavily constrained by experimental data
on the photon mass, see e.g.~\cite{Luo:2003rz,Goldhaber:2003dx}. It should 
also be emphasised that the Elko spinors are neutral with respect to $U(1)$ 
and $SU(n)$.

Nonetheless, such an interaction, though certainly very small, would be present
in nature if also the Elko spinors would turn out to be more than just a 
theoretical construct. Consequently, the photons would become massive and
the resulting field theory of a massive spin one field could consistently be
coupled to Einstein-Cartan theory. Unfortunately, this possibility does not 
contain a general scheme to couple gauge fields minimally to Einstein-Cartan 
theory. The simplest interaction term of Elko spinors with Yang-Mills 
fields\footnote{I thank Daniel Grumiller for elucidating this point.}
would be of the form
\begin{align}
      \lan_{\rm int}=c_{\rm FF}\, {\rm Tr} (F_{ab}^\mathcal{A} F^{ab}_\mathcal{A})\, 
      \dual{\lambda} \lambda \,,
      \label{eq:q3}
\end{align}
where with $\mathcal{A}$ the internal $SU(n)$ group index was denoted. As in 
the above mentioned case, the coupling constant $c_{\rm FF}$ must be extremely 
small.

\section{Summary and conclusions}
\label{sec:sum}

The dual helicity structure of Elko spinors motivated the study of Elko spinors 
in spacetimes with torsion. In contrast to the Dirac spinors, the Elko spinors'
spin angular momentum tensor shows a much richer structure. Physically this
can be explained from the mentioned dual helicity structure, whereas from a
mathematical point of view it is a consequence of the scalar-like Lagrangian 
with spinorial fields.

The Elko spinors are prime first principle candidates for dark matter. 
It was shown that the presence of Elko spinors would necessarily leads to 
massive photons. Hence, the resulting massive spin one field theory could 
be coupled consistently to the Einstein-Cartan theory.

In a series of papers~\cite{Griffiths:1980,Griffiths:1981ym,Griffiths:1982}, 
the extended (with torsion) spin-coefficient formalism was developed by
Griffiths and Jogia. It would be promising to apply this extended formalism
to the Einstein-Cartan-Dirac system so that the geometrical structure induced 
by Elko spinors in spacetimes with torsion can be analysed more invariantly.

In order to further study the physics of the Einstein-Cartan-Elko system,
cosmological investigations seem to be the most natural starting point,
since the number of degrees of freedom is greatly reduced. 
On the other hand, gravity theories with torsion have also 
been elaborately studied by several authors, see e.g.~\cite{Goenner:1984},
assuming the spacetime to be homogeneous and isotropic. 
If the cosmological principle is applied~\cite{Tsamparlis:1979}, the torsion 
(or contortion) tensor is restricted to the following non-vanishing components,
a vector and an axial vector torsion part. Therefore, since all technical 
ingredients are now known, the analysis of the cosmological Einstein-Cartan-Elko 
can in principle begin and will be the subject of further investigation.
 One of the most interesting issues will certainly be the study of torsion
effects~\cite{Kao:1993nb,Chatterjee:1993rc,Capozziello:2002kw,Boehmer:2005sw} 
in the inflationary epoch, however, driven by Elko spinors.

\acknowledgments
I would very much like to thank Dharam Ahluwalia-Khalilova for making this 
work possible, and Daniel Grumiller for the useful discussions. This project is 
supported by research grant BO 2530/1-1 of the German Research Foundation (DFG).

\end{document}